\documentclass[preprint,authoryear,12pt]{elsarticle}

\usepackage{slashbox}
\usepackage{setspace}  
\usepackage{hyperref}

\usepackage{amssymb}

\journal{JASIST}

\begin{document}

\begin{frontmatter}

\title{The relationship between acquaintanceship and coauthorship in scientific collaboration networks\footnote{This is a preprint. Published version can be found at DOI: 10.1002/asi.21629.}}

\author{Alberto Pepe}

\address{Center for Astrophysics \& Institute for Quantitative Social Science, Harvard University}

\begin{abstract}
This article examines the relationship between acquaintanceship and coauthorship patterns in a multi-disciplinary, multi-institutional, geographically distribu\-ted research center. Two social networks are constructed and compared: a network of coauthorship, representing how researchers write articles with one another, and a network of acquaintanceship, representing how those researchers know each other on a personal level, based on their responses to an online survey. Statistical analyses of the topology and community structure of these networks point to the importance of small-scale, local, personal networks predicated upon acquaintanceship for accomplishing collaborative work in scientific communities.
\end{abstract}


\end{frontmatter}

\section{Introduction}
Numerous way to study scientific collaboration exist. Scholarly \textit{coauthorship}---the joint publication of scholarly articles---is a widely used indicator of collaboration in scientific circles. In specialized literature, coauthorship networks have been employed to study collaboration patterns within entire scientific domains, e.g., neuroscience \citep{braun}, biomedicine \citep{newman:2000}, and nanoscience \citep{schummer:2004}, as well as within more specialized scientific circles, e.g., the communities of researchers involved in information visualization \citep{borner}, genetic programming \citep{tomassini}, and environmental wireless sensing \citep{pepe:scientometrics2010}. 

Besides coauthorship, studies of scientific collaboration have also covered the importance of \textit{acquaintanceship} for scientific communication, labor management and coordination. Studies of this kind employ ethnographic methods such as direct interviews, participant observations of meetings, and sociometric surveys to explore the role of personal networks in modeling scientific advancement. Early studies of acquaintanceship in science have analyzed the impact of task-based workflows \citep{kraut:1987}, face-to-face communication \citep{lievrouw:1987}, and scientific meetings \citep{liberman:1997} on the structure of interpersonal ties and knowledge flows among scientists in various scientific domains. More recently, scientific acquaintanceship studies have touched upon emerging cyber-infrastructure initiatives, by exploring the role of formal and informal personal knowledge exchange in the context of ``virtual science'' laboratories \citep{george-chin:2002} and multi-institutional interdisciplinary research centers \citep{noriko:2003}. 

While the literature covers many aspects of both coauthorship and acquaintanceship in the context of scientific communities, rarely has their relationship been examined in detail. This is because collecting data about acquaintanceship and related social traces is a time-consuming, elaborate and error-prone procedure, especially for large scientific communities. Outside of science, with growing availability of online corpora containing social indicators, it is possible to employ large datasets harvested from the web to infer acquaintanceship ties. Personal connections as variegated as business, friendship and romantic connections are increasingly becoming available online and are being used for large-scale social network analyses. Examples include: a comparison of online and offline friendship ties \citep{tong_fb}, an analysis of acquaintanceship ties and geographic coincidence \citep{Crandall:2010uq}, and the structural and temporal evolution of a dating website community \citep{dating:2004}. But a similar online footprint of acquaintanceship ties is not available for most scientific communities.

For this reason, coauthorship networks are sometimes used as proxies for acquaintanceship networks. For example, in \textit{Who is the best connected scientist?} \citep{newman:2000}, Newman constructs a coauthorship network and employs it as an acquaintanceship network, assuming that ``it is probably fair to say that most people who have written a paper together are genuinely acquainted with one another'' \citep[p. 339]{newman:2000}. Is this a reasonable assumption? Understanding the relationship that exists between coauthors and acquaintances is important not only to elucidate the social component that exists at the basis of modern scientific collaborations, but also to measure the extent by which scholarly coauthorship networks can be used as surrogates for social networks. In traditional research settings, it was very likely that coauthors of a scholarly article were acquainted with each other at a personal level. But in recent times, scientific research is increasingly being conducted in centers which can be extremely large, variegated in their disciplinary component, and geographically distributed over different cities, countries and continents. For example, many ongoing collaborations in physics and astrophysics are so large that some publications include hundreds of authors. For these articles, it is virtually impossible to discern the nature and extent of individual contributions to research, let alone derive acquaintanceship patterns \citep{solomon:2009}. Similarly, many modern cyber-infrastructure enterprises, in the form of collaboratories and e-science centers, are so variegated in their disciplinary and geographical configuration that their researchers rely predominantly upon computer supported technologies for their communication, collaboration and organization \citep{olson,Bos:2007fk,finholt_collab}. How are coauthorship and acquaintanceship patterns related to each other in such emerging scientific settings? This article addresses this question by a comparative network analysis of collaboration in a distributed, multi-disciplinary research environment.

\section{Study, data and instruments}
\subsection{Motivation and research questions}
The scientific community analyzed here is pivoted around the Center for Embedded Networked Sensing (CENS), a National Science Foundation Science and Technology Center established in 2002, involved in the development and application of sensor network systems to critical scientific and societal pursuits. CENS features many of the characteristics of modern ``collaboratories'': it is multi-institutional, multi-disciplinary, and geographically distributed. It includes five member universities in California: University of California, Los Angeles, University of Southern California, University of California, Riverside, California Institute of Technology, and University of California, Merced. It includes 300 faculty, students, and staff specialized in disparate academic disciplines, ranging from computer science to biology and environmental science, with additional partners in arts, architecture, and public health. The type of research conducted at CENS spans a wide spectrum of disciplines and applications requiring continuous cooperation among individuals who, otherwise, would probably not interact beyond the walls of traditional university departments and faculties. CENS features a headquarter base located at UCLA, yet CENS-related work is conducted in departments, labs, and remote field locations situated near all five member institutions as well as at partner institutions in the U.S. and abroad. These institutions (and sometimes even departments) are sufficiently distant from one another to prevent continuous physical interactions among scientists: computer-supported communication is at the basis of their collaborative work. 

This variegated institutional, disciplinary, and geographical arrangement makes CENS a convenient environment to investigate scientific collaboration in modern research settings. Two manifestations of collaboration among CENS researchers are investigated. The first is \textit{coauthorship}, defined as the joint authoring of scholarly artifacts. In a coauthorship network, two individuals are linked to one another if they are coauthors. The second manifestation is \textit{acquaintanceship}, defined as a relationship of personal knowledge characterized by mutual face and name recognition. In an acquaintanceship network, two individuals are linked if they indicate to recognize and know each other at a personal level. These two networks are analyzed and compared. To begin with, it is interesting to explore the overarching statistical properties of the networks of acquaintanceship and coauthorship to understand whether they differ in a systematic way. Do they have a similar configuration? The first portion of this research is aimed at exploring the mechanisms by which CENS researchers write papers and make acquaintances with one another. It can be summarized as follows: 

\begin{quote}
\textbf{RQ1.} What is the topology of the coauthorship and acquaintanceship networks in distributed, multi-disciplinary research environments? What is the procedure by which researchers form ties with their coauthors and acquaintances?
\end{quote}

The second portion of this study presents a comparative analysis of the coauthorship and acquaintanceship networks. It is clear that in collaboratories like CENS, computer supported technologies are enabling new modes of science, scientific advancement and collaboration. Yet, we still have to fully understand how these emerging forms of remote collaboration function in the absence of interpersonal knowledge and contact, or similarly, when such contact is predominantly mediated via digital networks. It has been recently noted, for example, that physical distance between researchers that work across organization boundaries is the most likely factor to hinder scientific collaboration and coordination \citep{cummings}. What is the proportion of CENS researchers that produce joint scholarly work but have never met each other in person? In this part of the study, a comparative network analysis is conducted to understand how well communities of coauthors and acquaintances are representative of one another. With these notions in mind, the second guiding question can be summarized as follows:

\begin{quote}
\textbf{RQ2.} Is coauthorship a suitable proxy for acquaintanceship in multi-institutional, distributed, scientific communities? How do communities of coauthors and acquaintances overlap?
\end{quote}

Findings from this research are presented in the next sections. In the remainder of this section, I discuss the data collection procedures and the instruments employed in this study.

\subsection{Bibliographic data and coauthorship network}
\label{sec4:bibliographic}
The coauthorship network of this scientific community was constructed from its bibliographic record, gathered from all scholarly items listed in seven available CENS Annual Reports (2003--2009). A publication database was assembled, consisting of 608 papers published by a total of 391 unique individuals over a period of ten years (2000--2009). Table \ref{tab:pub_stats} summarizes the distribution of papers analyzed by publication type, and number of authors. Roughly two-thirds of publications analyzed are papers in conference proceedings, while journal articles take up the other third of the volume of publications. The distribution of items per number of authors reveals that about half of all publications are authored by two or three individuals. Author lists rarely exceed six authors. 

\begin{table}[]
\caption{\label{tab:pub_stats}Basic statistics for the collected bibliographic data: paper distribution by publication type, publication year, and number of authors}
\centering
\begin{tabular*}{\textwidth}{@{\extracolsep{\fill}}lr}
\hline
\textbf{Paper type}&\textbf{$n$ = 608}\\
\hline\hline
Conference proceedings&400\\
Journal article&189\\
Book chapter&18\\
Book&1\\

\hline
\textbf{Number of authors}&\textbf{$n$ = 608}\\
\hline\hline
1&59\\
2&155\\
3&158\\
4&94\\
5&59\\
6&32\\
7&19\\
8&13\\
9&5\\
10+ (where 14 is the maximum number of authors found)&14\\
\hline\hline
\end{tabular*}
\end{table}

This bibliographic record was employed to construct the coauthorship network---a network in which nodes represent authors and edges represent the extent of coauthorship activity. The network is weighted and the edge weights are established by partitioning a set value for every publication. In order to determine the weights between nodes, i.e., the strength of collaboration among coauthors, I use a weighting mechanism by which the weight of the edge between nodes $i$ and $j$ is:
\begin{equation}
\label{formula:newman}
w_{ij} = \displaystyle\sum_{k} \frac{\delta^k_i \delta^k_j}{n_k-1},
\end{equation}
where $\delta^k_i$ is $1$ if author $i$ collaborated on paper $k$ (and zero otherwise) and $n_k$ is the number of coauthors of paper $k$. As such, this weighting mechanism confers more weight to small and frequent collaborations, based on the assumptions that: i) publications authored by a small number of individuals involve stronger interpersonal collaboration than multi-authored publications, and ii) authors that have authored multiple papers together know each other better on average and thus collaborate more strongly than occasional coauthors. Using this weighting mechanism, a network of coauthorship is constructed. It is depicted and analyzed in the next section.

\subsection{Survey instrument and acquaintanceship network}
A survey instrument was designed and implemented to ask CENS coauthors to indicate their acquaintances. This study is specifically aimed at investigating a form of personal acquaintanceship that involves name and face recognition. For this reason, surveyed subjects are asked to indicate as acquaintances individuals that \textit{a)} they have previously met in person, and that \textit{b)} they would say ``hi'' to if they bumped into them. This description is aimed at discriminating between forms of genuine personal acquaintanceship, and other forms of acquaintanceship that do not necessarily involve a personal component, e.g., name cognizance based solely on email correspondence.

The survey roster consisted of 388 individuals, i.e., all the authors extracted from the bibliographic record (391) minus the individuals for whom no contact information could be found (3). Individuals in the roster were invited to take part in the survey via a recruitment letter, sent via electronic mail. The survey instrument was structured in two parts. 

In the first part, respondents are asked to select their acquaintances from a list of all of the individuals in the survey roster. Individuals are identified by name and a thumbnail picture. In order to aid recognition, individuals are grouped together by department and institutional affiliation. Respondents select their acquaintances on this page and after submitting the data are taken to the second part of the survey. In the second part, they are then asked to indicate the nature and length of the relationship with their acquaintances. Respondents are prompted with the list of people that they selected as acquaintances in the previous page. For each individual they indicate as acquaintance, two questions are asked: \textit{When did you first meet?}  (2001 or earlier, 2002, 2003, 2004, 2005, 2006, 2007, 2008, This year), and \textit{How often are you in touch?} (At least once/week, At least once/month, Occasionally, Rarely or never). Screen-shots of the first and second parts of the questionnaire are included in the Appendix A, Figures \ref{fig:q1} and \ref{fig:q2}.

A total of 191 responses were collected. Some basic statistics relative to the data collected via the survey instrument are summarized in Table \ref{tab:survey_stats}. Nearly half of respondents invited to fill in the survey (49\%) participated in the study. The rest were either not reachable (4\%), or did not respond to the survey by the end of data collection (47\%). About one third of respondents (39\%) only completed the first part of the survey. The vast majority of respondents indicated a number of acquaintances ranging between 5 and 40. 

\begin{table}[]
\caption{\label{tab:survey_stats}Basic statistics for the collected social survey data.}
\centering
\begin{tabular*}{\textwidth}{@{\extracolsep{\fill}}lr}
\hline
\textbf{Survey response}&$n = 388$\\
\hline\hline
Respondents&191 (49 \%)\\
Non-respondents&182 (47 \%)\\
Unreachable&15 (4 \%)\\
\hline
\textbf{Portion of survey completed}&$n = 191$\\
\hline\hline
Full survey&116 (61 \%)\\
Only part one&75 (39 \%)\\
\hline
\textbf{Number of acquaintances}&$n = 191$\\
\hline\hline
1-5&12\\
5-10&21\\
10-20&47\\
20-30&38\\
30-40&19\\
40-50&12\\
50-60&16\\
60-70&9\\
70-100&10\\
100+ (maximum is 197)&7\\
\hline\hline
\end{tabular*}
\end{table}

\begin{figure}[]
\centering
\includegraphics[width=0.8\textwidth]{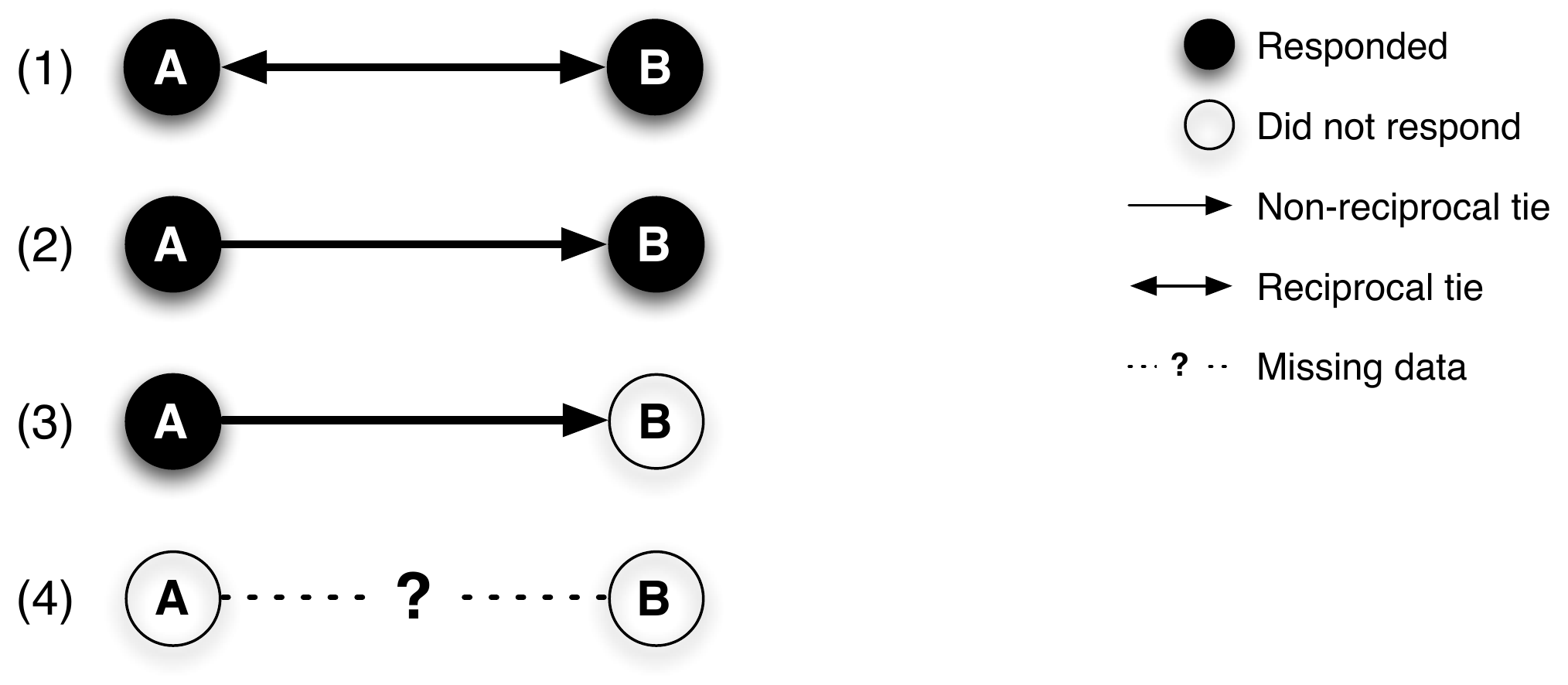}
\caption{Four possible classes of acquaintanceship ties:  (1) complete data, reciprocal tie, (2) complete data, non-reciprocal tie, (3) incomplete data, non-reciprocal tie, and (4) missing data. Filled (black) nodes represent individuals who took the survey, while blank (white) nodes depict individuals who did not respond.}
\label{fig:four_typology} 
\end{figure}

Responses to the survey were obtained independently from one another and resulted in either reciprocal or non-reciprocal ties, as well as complete or incomplete data. Four different scenarios based on the responses obtained are summarized in Figure \ref{fig:four_typology}:  (1) complete data, reciprocal tie, (2) complete data, non-reciprocal tie, (3) incomplete data, non-reciprocal tie, and (4) missing data.

\renewcommand{\baselinestretch}{1}
\renewcommand{\baselinestretch}{2}

The typology presented in Figure \ref{fig:four_typology} includes both reciprocal ties (represented in a directed network by bidirectional edges) and non-reciprocal ties (unidirectional edges). For the purpose of this study, network directionality is not of fundamental importance, mostly because the coauthorship network is natively undirected and maintaining information about acquaintanceship reciprocity would not enable additional analyses and comparisons. For this reason, directed acquaintanceship ties are converted to undirected edges using a an available-case analysis \citep{little_rubin}, i.e., by including all reciprocal and non-reciprocal ties with both complete and partial descriptions. Thus, the acquaintanceship network includes ties (1), (2) and (3)\footnote{In order to justify the reconstruction of case (2) and (3) ties, two tests were performed. The first test involved checking that the population of respondents was not systematically different from the population at large, i.e., that the distribution of respondents in terms of departmental and institutional affiliation matched that of the entire population. The second test involved manually inspecting incomplete data.}, from Figure \ref{fig:four_typology}, broken down as presented in Table \ref{tab:survey_res_cases}. 

\renewcommand{\baselinestretch}{1}
\begin{table}[]
\caption{\label{tab:survey_res_cases}A summary of the data collected in the online survey of acquaintanceship.}
\centering
\begin{tabular*}{\textwidth}{@{\extracolsep{\fill}}crrr}
\hline
Tie case&Data&Type of tie&\# ties\\
\hline\hline
(1)&Complete&Reciprocal&\textbf{2756} (45\%)\\
(2)&Complete&Non-reciprocal&\textbf{976} (16 \%)\\
(3)&Incomplete&Non-reciprocal&\textbf{2451} (39 \%)\\
\hline
Totals&&&\textbf{6183} (100\%)\\
\hline\hline
\end{tabular*}
\end{table}
\renewcommand{\baselinestretch}{2}

In the second part of the survey, respondents were asked to indicate how long they have known their acquaintances for and how frequently they are in touch with them. Particularly relevant to this discussion is the data regarding the frequency of communication, that was used to compute edge weight. Table \ref{tab:survey_res_summary2} summarizes the data collected in this portion of the study.

\renewcommand{\baselinestretch}{1}
\begin{table}[]
\caption{\label{tab:survey_res_summary2}A summary of the data collected in the second part of the online survey of acquaintanceship.}
\centering
\begin{tabular*}{\textwidth}{@{\extracolsep{\fill}}lr}
\hline
\textit{How often do you communicate with [name]?}&\\
\hline\hline
Did not respond (no data available)&1668 (26\%)\\
Responded:&4621 (74\%)\\
\textit{..........Rarely or never (0.25)}&2245\\
\textit{..........Occasionally (0.50)}&1539\\
\textit{..........At least once/month (0.75)}&421\\
\textit{..........At least once/week (1.0)}&416\\
\hline\hline
\end{tabular*}
\end{table}
\renewcommand{\baselinestretch}{2}

As shown in Table \ref{tab:survey_res_summary2}, respondents provided data relative to the frequency of communication for about three quarters of the total number of acquaintances indicated. Moreover, a quick analysis of the distribution of responses reveals that nearly half of all acquaintances (2245 out of 4621) communicate rarely or never. About a third of all ties (1539 out of 4621) are based on occasional communication. Only about a fifth of all ties relies on frequent communications (once a month and once a week). This information is used to assign a weight to the edges connecting acquaintances. Frequent communication are given a higher weight, based on the assumption that frequent communication involves a higher degree of cognizance among individuals, regardless of the nature of the relationship (formal or informal). If respondents indicate to communicate once a week, the edge among them is weighted $1.0$; acquaintances communicating at least once a month are assigned with a weight of $0.75$. Occasional communication is weighted $0.5$ and even less frequent communication is weighted $0.25$. All collected data (including partial and non-reciprocal data) were documented and used construct weighted edges. The resulting weighted network of acquaintanceship is presented and depicted in the next section.

\section{Analysis and Results}
\subsection{Statistical properties and topology of the networks}
Some statistical properties of the constructed coauthorship and acquaintanceship networks are presented in Table \ref{tab:netstats_fin}. An analysis of these statistics provides insights into the collaborative configuration of this scientific community and its underlying mechanisms.

\begin{table}[h!]
\caption{\label{tab:netstats_fin}Statistical properties of the studied coauthorship and acquaintanceship networks.}
\centering
\begin{tabular*}{\textwidth}{@{\extracolsep{\fill}}lrr}
\hline
Property&Coauthorship&Acquaintanceship\\
\hline \hline
Nodes, $n$&391&385\\
Edges, $m$&$1\,747$&$4\,805$\\
Average path length, $\ell$&2.952&2.427\\
Clustering coefficient, $C$&0.301&0.359\\
Degree assortativity, $r$&0.013&-0.060\\
\hline \hline
\end{tabular*}
\end{table}

The coauthorship network includes a total of 1747 scholarly collaborations (edges) among 391 authors (nodes). The acquaintanceship network, with 385 nodes, is comparable in size to the coauthorship network: the results of the survey covered almost the entire range of coauthors. What differs greatly, however, is the number of edges in the acquaintanceship network, which are far more numerous than are coauthoring ties (4805 vs. 1747). This indicates that, at a broad level, there are more researchers connected to each other via interpersonal relationships than by coauthorship. This simple finding is a first probe into the first research question proposed above (RQ.1). It portrays a collaborative ecology in which acquaintanceship ties are far more frequent than coauthoring relationships, despite CENS, like most collaboratories, being built upon the underlying assumption of distributed, remote collaboration (which does not necessarily require a personal level of acquaintanceship).

Two other statistical properties indicated in Table \ref{tab:netstats_fin} are the average path length and the clustering coefficient. The \textit{average path length} is the average number of steps needed to connect any two nodes in a network. Its value is relatively small across the two networks: it takes on average two to three steps to connect any two individuals both in the coauthorship and acquaintanceship networks, connoting well-connected networks in which information can transfer easily between nodes. The \textit{clustering coefficient}, i.e., the density of cliques in a network, is also similar in both networks and is relatively low, indicating that both networks are not significantly dense. The fact that both networks exhibit similar clustering coefficients suggests that the modalities by which researchers connect do not change significantly based on the platform of collaboration: whether it is scholarly papers or interpersonal knowledge relationships, researchers form communities that result in similar topologies. Many real networks, including scientific coauthorship and acquaintanceship networks, have been observed to be \textit{small-world} networks \citep{watts_nature_98}, i.e., they have low average path length and high clustering coefficient. The networks presented here slightly deviate from this small-world model\footnote{For a comparison, consider the coauthorship networks in biology ($\ell = 4.92$, $C = 0.60$) \citep{newman2}, and neuroscience ($\ell = 5.7$, $C = 0.76$) \citep{barabasi2002}, and acquaintanceship networks of movie actors ($\ell = 3.48$ and $C = 0.78$) \citep{watts_nature_98} and corporate company directors ($\ell = 4.60$ and $C = 0.88$) \citep{davis:2003}.}. 

A final statistical property calculated on the CENS networks of coauthors and acquaintances is assortativity \citep{newman:assort}. \textit{Assortativity}, is a measure of homophily in a network, i.e., it measures the tendency for nodes with similar characteristics to attach to one another. Assortativity can be calculated for a numer of discrete or categorical features of a network. For example, a measure of assortativity by ethnicity in a social network would reveal how individuals of same ethnicity preferentially attach to one another. A widely studied measure of assortativity in scientific networks is degree assortativity, i.e., the tendency for scientists to preferentially attach to others with similar network degree, where degree is simply the number of edges attached to a node. Thus, a high assortativity coefficient in a coauthorship network ($r\to1$) means that prolific authors collaborate preferentially with other prolific authors. Inversely, a low assortativity coefficient ($r\to0$) indicates that prolific authors collaborate both with prolific and non-prolific authors, without preference. Network studies of scientific coauthorship and acquaintanceship have shown moderate degree assortativity coefficients\footnote{Examples are the networks of coauthorship in physics ($r = 0.363$) \citep{newman2}, biology ($r = 0.127$) \citep{newman2}, and mathematics ($r = 0.120$) \citep{erdos,Grossman95}. These results indicate that prolific authors in these disciplines tend to coauthor papers with other physicists of similar high standing. Similarly, social networks of movie actors \citep{watts_nature_98} and company directors \citep{davis:2003} have both been found to exhibit some form of assortativity ($r = 0.208$ and $r = 0.276$, respectively). This means that both the world of entertainment and that of corporations are ecologies in which prestige and popularity matter.}. The near-zero assortativity coefficients of the coauthorship ($r = 0.013$) and acquaintanceship networks ($r = -0.060$) of CENS, indicate a contrasting scenario: there is no significant tendency for individuals to write papers or establish friendships with others with a similar number of collaborators. 

The statistical properties of Table \ref{tab:netstats_fin} can be used to draw some preliminary conclusions regarding the overall topology of coauthorship and acquaintanceship patterns at CENS (RQ.1). As discussed above, I find that acquaintanceship ties are far more frequent than coauthoring ties. This result, alone, highlights the prevalence of personal-level relationships in a distributed collaborative environment. From a topological perspective, both coauthorship and acquaintanceship networks are comparable: they feature short average path length, low clustering, and near-zero degree assortativity. These results point to a well-connected, fluid, and inclusive ecology of collaboration. This configuration, in turn, indicates that no specific ``attachment'' mechanisms can be identified for the formation of coauthoring and acquaintanceship ties: researchers with different network degrees (e.g., both prolific and marginal authors) ``attach'' to one another without abiding to definite preferential rules. This result is further discussed in the next section. 

\subsection{Comparison of the networks and community structure}
The social networks studied in this article portray two different relationships (coauthorship and acquaintanceship) among the same set of individuals (CENS researchers). A simple way to determine how these networks are related to one another is by calculating the portion of coauthorship ties that are also acquaintanceship ties, and, thus, the number of individuals who are both coauthors and acquaintances. This breakdown is provided in Table \ref{tab:acq_non}. The results in the table show that the vast majority of coauthors (over three quarters) also know each other on a personal level. Also, most coauthoring events in this scientific community (about two thirds) were performed with acquaintances.

\begin{table}[!]
\caption{\label{tab:acq_non}Acquainted and non-acquainted coauthors.}
\centering
\begin{tabular*}{\textwidth}{@{\extracolsep{\fill}}lll}
\hline
Quantity&Nodes&Edges\\
\hline \hline
Acquainted coauthors&305 (78\%)&$1\,134$ (64\%)\\
Non-acquainted coauthors&86 (22\%)&613 (36\%)\\
\hline
Total coauthors&391&$1,747$\\
\hline \hline
\end{tabular*}
\end{table}

A more precise statistical measure of inter-network comparison can be obtained by performing a Quadratic Assignment Procedure (QAP) correlation \citep{krack}. Given two networks that represent different relationships (edges) of the same individuals (nodes), QAP methods calculate the Pearson's correlation coefficient on corresponding cells. In the case considered here, a QAP correlation reveals the probability that a coauthorship tie among two individuals is related to the probability of an acquaintanceship tie among the same individuals. In other words, are coauthors also likely to be acquaintances? The observed QAP correlation between the coauthorship and acquaintanceship networks was found to be $r = 0.372$ ($p < .001$) supporting the proposition that individuals who coauthor papers are likely to know each other in person\footnote{QAP correlation tests between the coauthorship network and a randomly generated network of the same size returned $r = -.002$ ($p = 0.719$)}. These findings indicate that in the context of the medium-sized, inter-institutional, distributed collaboration analyzed here, coauthors are likely to be acquainted with one another (RQ.2). This result is further discussed in the next section. 

While the comparative analysis above provides an understanding of how these networks overlap at a broad level, further analysis is required to understand more in detail the actual arrangement of collaboration circles within these networks. A study of community structure enables a network to be partitioned into clusters and comparatively analyze how different configurations of collaboration patterns overlap with each other. \textit{Community structures}, also called structural communities, are ``cliquish'' groupings of nodes that are highly connected between them, but poorly connected to other nodes \citep{girvan-2002}. In the coauthorship and acquaintanceship networks, the clusters detected via a structural analysis correspond to communities of collaborating researchers that write papers together and know each other, respectively. How do these clusters compare with each other?

\begin{figure}[h!]
\centering
\includegraphics[width=0.85\textwidth]{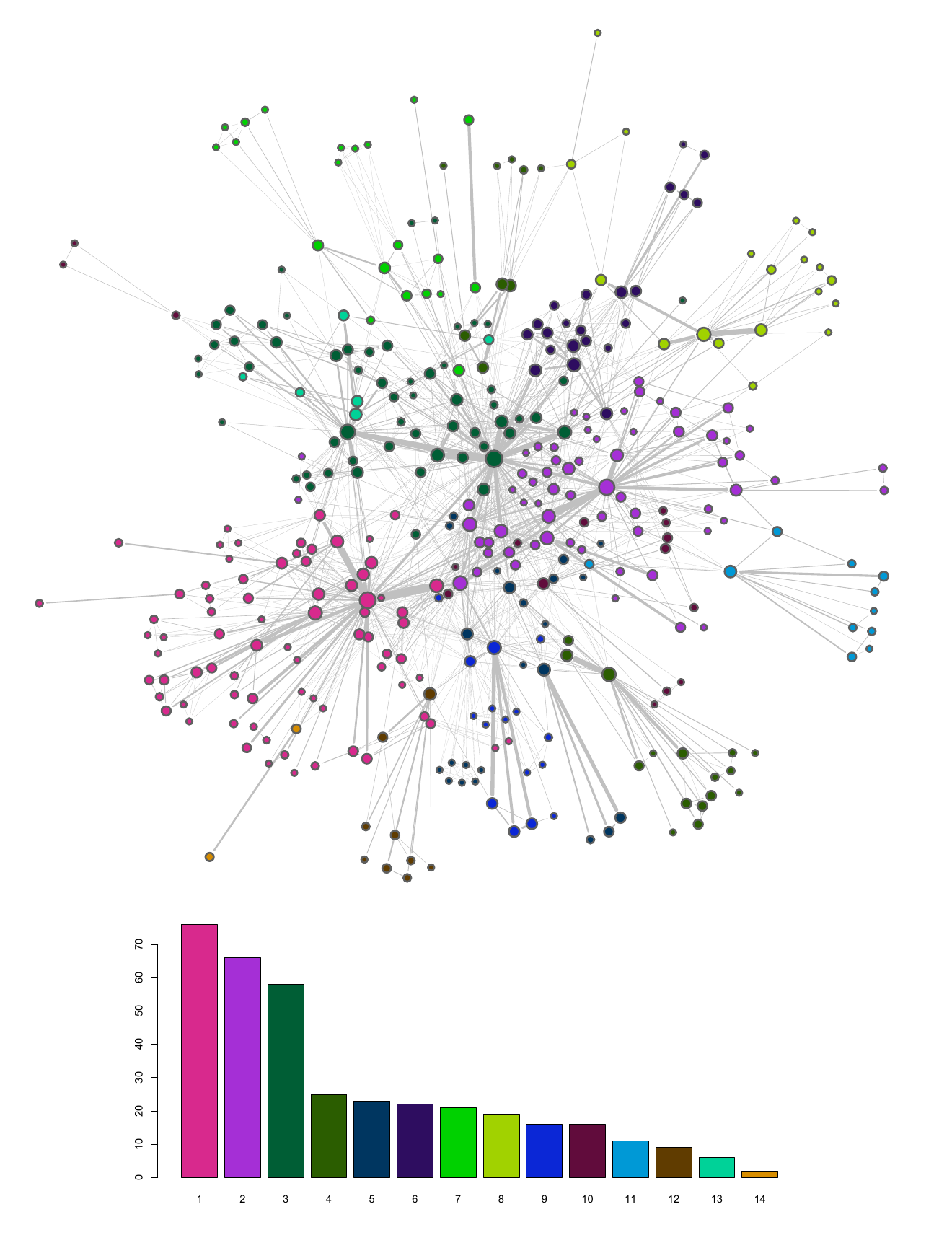}
\caption{Structural communities in the coauthorship network detected according to the spinglass algorithm. Node color represents structural community membership. Node diameter represents betweenness centrality. Associated histogram describes the frequency distribution of each community. }
\label{fig:coauth_communities} 
\end{figure}

\begin{figure}[h!]
\centering
\includegraphics[width=0.85\textwidth]{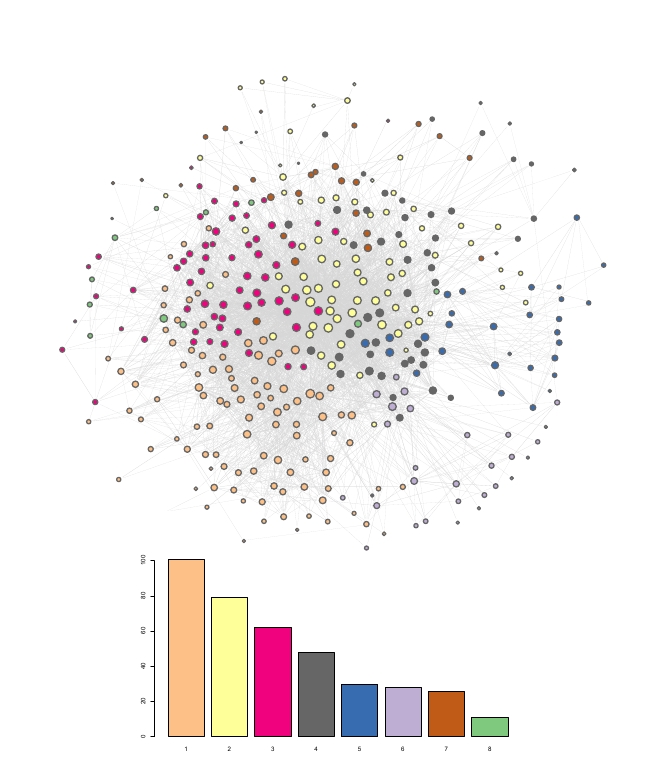}
\caption{Structural communities in the acquaintanceship network detected according to the spinglass algorithm. Node color represents structural community membership. Node diameter represents betweenness centrality. Associated histogram describes the frequency distribution of each community. }
\label{fig:acq_communities} 
\end{figure}

The community detection method used here is the spinglass algorithm \citep{spinglass:reichardt2006}. This method relies on an analogy between the statistical mechanics of networks and physical spin glass models to deconstruct a network into communities. In doing so, it assigns a community membership value to each node. Thus, individuals who are in the same structural community are given the same membership value. The structural communities found in the collaboration networks via the spinglass algorithm are diagrammed in Figures \ref{fig:coauth_communities} and \ref{fig:acq_communities}. Each figure presents a network with nodes colored according to the structural community that they belong to. Each community is represented using a different color (or shade). Node diameter represents the betweenness centrality score of nodes, where more central nodes have larger diameters. The histogram associated with each Figure describes the frequency distribution of each community, i.e., the number of scholars in each identified structural community. 

\renewcommand{\baselinestretch}{1}
\renewcommand{\baselinestretch}{2}

\renewcommand{\baselinestretch}{1}
\renewcommand{\baselinestretch}{2}

A total of $14$ structural communities were found in the coauthorship network (Figure \ref{fig:coauth_communities}). The population distribution histogram shows that three single communities (with populations 76, 66, and 58) cover about half of the entire coauthorship population. The remaining coauthorship communities are smaller in size (with an average of 15 members). The acquaintanceship network was partitioned into $8$ communities (Figure \ref{fig:acq_communities}). The population distribution shows that there are three acquaintanceship communities that are highly populated (with populations 101, 79, and 62) and the remainder of the nodes more or less evenly distributed in the remaining five communities. 

The network diagrams of Figures \ref{fig:coauth_communities} and \ref{fig:acq_communities} show the repartition of the networks into clusters of collaboration. More analysis is needed to reveal how these arrangements of collaboration differ with each other, i.e., how researchers group into communities of coauthorship and acquaintanceship. A statistical comparison to measure community overlap in the networks is performed by use of tests of independence that determine whether community membership in one network is dependent or independent on membership in the other network. A contingency table portraying a numerical representation of the overlap between community membership in these two networks is presented in Table \ref{tab:coauth-acq-overlap}. This table displays the association between the coauthorship and acquaintanceship networks, i.e. between circles of collaboration depicted in Figures \ref{fig:coauth_communities} and \ref{fig:acq_communities}. Columns in this table (the $x$-axis) list community membership values in the acquaintanceship network and rows ($y$-axis) in the coauthorship network. 

\renewcommand{\baselinestretch}{1}
\begin{table}[h!]
\caption{\label{tab:coauth-acq-overlap} Contingency table presenting the community membership association between the coauthorship (``C'', columns) and acquaintanceship (``A'', rows) networks with total membership counts (``T'').}
\begin{footnotesize}
\centering
\begin{tabular*}{\textwidth}{@{\extracolsep{\fill}}c|rrrrrrrrrrrrrr|r}
\backslashbox{A}{C}&1&2&3&4&5&6&7&8&9&10&11&12&13&14&T \\
\hline
1&50&7&37&&&&1&2&2&&&&&2&101\\
2&1&23&10&8&8&1&18&3&1&&1&&&&74\\
3&&34&5&&&1&&14&1&&1&&&&56\\
4&6&2&2&1&&&1&&12&15&&&6&&45\\
5&1&&&16&13&&&&&&&&&&30\\
6&18&&&&&&&&&&&9&&&27\\
7&&&4&&1&20&&&&1&&&&&26\\
8&&&&&1&&1&&&&9&&&&11\\
\hline
\textbf{T}&76&66&58&25&23&22&21&19&16&16&11&9&6&2&370\\
\end{tabular*}
\end{footnotesize}
\end{table}
\renewcommand{\baselinestretch}{2}

\begin{figure}[h!]
\centering
\includegraphics[width=1.0\textwidth]{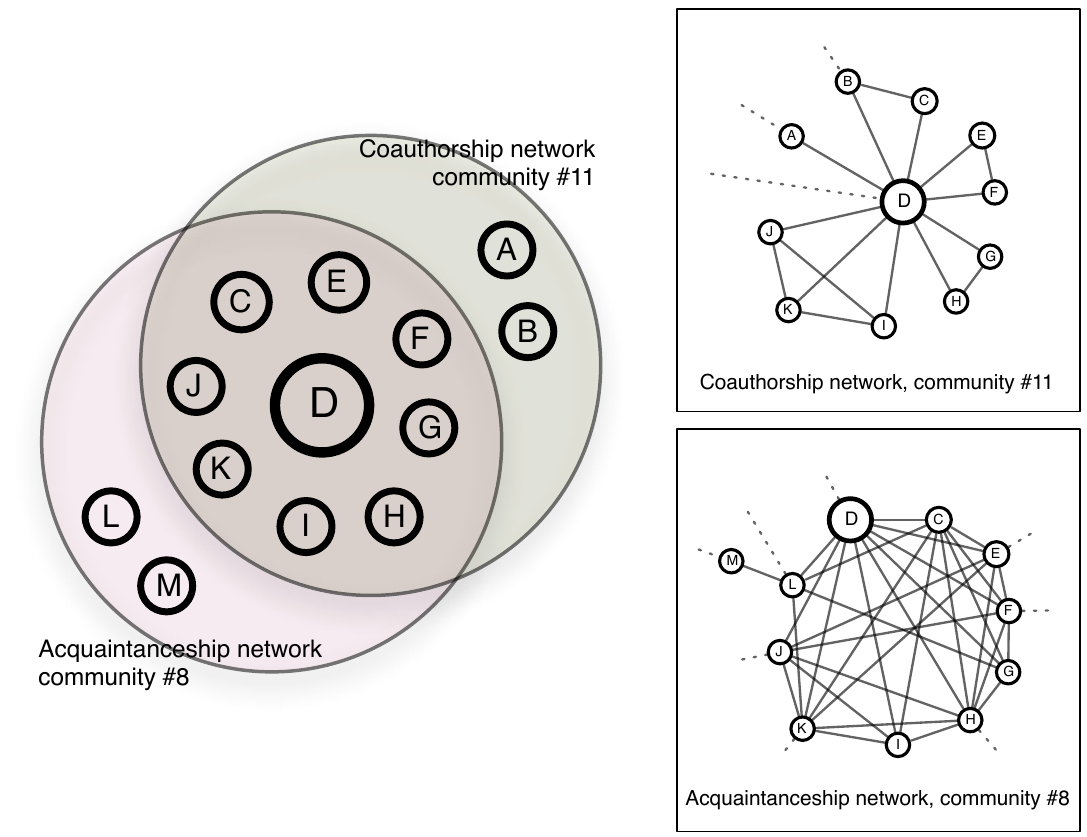}
\caption{Anecdotal example: a case of overlap of coauthorship and acquaintanceship communities. The inset image in the top right corner shows community \# 11 in the coauthorship network ($n = 11, m = 16$). The inset image at the bottom right corner shows community \# 8 in the acquaintanceship network ($n = 11, m = 29$). The main image shows the overlap between these two communities.}
\label{fig:soatto_overlap} 
\end{figure}

Analyzing the composition of Table \ref{tab:coauth-acq-overlap}, it can be seen that some communities overlap nearly perfectly, while others are highly partitioned. For example, community \# 11 in the coauthorship network and community \# 8 in the acquaintanceship network overlap almost completely. Both communities are composed of 11 individuals and 9 of them are found to be part of the same communities of coauthorship and acquaintanceship. Figure \ref{fig:soatto_overlap} depicts such anecdotal example. The figure shows that this coauthorship community is tightly centered around one individual (node $D$) whom collaborates separately with different groups (e.g., with $B$ and $C$, with $I$, $J$, and $K$, etc.). It is interesting to note that this community is very marginalized from the rest of the coauthorship network: only nodes $A$, $B$, and $D$ have connections with nodes outside this community (dashed edges). The tightly overlapping acquaintanceship community is composed of roughly the same nodes, but the relationships among them are more frequent and dense: nearly everyone is connected to everyone else in this acquaintanceship community. This indicates that even though not all members of a coauthoring community collaborate directly with each other, they are likely to know each other. Also, while the members of this community are separated from the rest of the coauthorship network (i.e., they only write papers with each other), they are considerably more integrated with the social network (dashed lines represent social relationships with the outside). 

\renewcommand{\baselinestretch}{1}
\renewcommand{\baselinestretch}{2}

While this anecdotal example provides a precise understanding of the scholarly and social organization of a given community, a significance test is needed to determine statistically the level of independence between community membership at a broader level. A reliable test of independence when dealing with samples of sufficiently large size that do not suffer from data sparseness is Pearson's $\chi^2$ \citep{reiser}. To overcome data sparseness, I directly manipulate data in Table \ref{tab:coauth-acq-overlap} by removing rows and columns with low frequency counts until less than one fifth of its cell expectations are below $5$, as suggested by \citet{cochran}. The reduced contingency table, subjected to independence test returned a high Pearson's $\chi^2$  score of $301.45$ ($p < .001$), indicating a strong, statistically significant correlation between community membership in these two networks\footnote{A Fisher's exact test on the same table also returned a p-value below $0.001$}. These results suggest that communities of coauthors and acquaintances overlap very well, i.e., coauthors of scholarly papers tend to form communities of collaboration that coincide with communities of personal acquaintances (RQ.2). 

\section{Discussion}
As scientific research endeavors become larger, more variegated in their disciplinary component, and more distributed geographically, the practices of scientific collaboration and communication are subject to radical changes. The disciplinary, institutional, and geographical diversity that characterizes emerging research endeavors such as collaboratories and cyberinfrastructure initiatives is such that researchers necessarily rely on computer supported technologies for their scientific collaboration, communication, and organization. Understanding the role that physical interaction and interpersonal communication play on distributed scientific workflows of this kind can prove useful in guiding, promoting and sustaining scientific collaboration \citep{Bos:2007fk}.

The research presented here elucidates the relationship between coauthorship and acquaintanceship patterns in a medium-sized distributed, multi-disciplinary collaboratory. The first portion of this research aims at unveiling the topological properties of these networks and the mechanisms by which they are generated. Results show that acquaintanceship relations are far more numerous than coauthorship ties: researchers know each other well within their coauthorship circle and beyond. From a topological perspective, however, the two networks are not very dissimilar: they share low average path length, and relatively low clustering coefficient. An analysis of degree assortativity reveals that both the coauthorship and acquaintanceship networks are fluid and inclusive: individuals of all degrees collaborate and know each other regardless of standing (near-zero degree assortativity coefficient). Overall, results of this topological analysis are telling of the social practices and authorship conventions of this specific community. The overwhelming predominance of acquaintanceship ties and the lack of preferential attachment rules substantiate the notion that the center under study is a fluid social ecology in which befriending patterns are ubiquitous and are not influenced by standing or prestige. In other words, I find no prestige-based dynamics by which prolific, well-known researchers collaborate with researchers of similar status. The near-zero degree assortativity suggests that as new members enter the CENS collaborative ecology (with low degree) they begin to collaborate both with prolific and with marginal individuals in their coauthorship and friendship networks. This constant mixing between individuals of all degrees keeps the network from becoming cliquish. This open social configuration not only benefits young students and newcomers, but it also contributes to making it a more diverse and inclusive collaborative environment. Similarly, the authorship of scholarly articles is an inclusive practice: scholars of different network centrality (both prolific and non-prolific authors) are observed to collaborate with each other without specific prestige-based attachment rules. This is not an isolated finding of this study; it is consonant with the authorship norms that operate at CENS and that support inclusion of scholars of all ranks and of many career levels in the preparation and publication of articles. These results imply that novel scholarly reward structures and recognition mechanisms may be emerging from cyberinfrastructure-enabled initiatives.

The second portion of this research aims to analyze the structural composition of the coauthorship and acquaintanceship networks in a comparative manner. Results of a QAP correlation test indicate that the networks overlap significantly and, thus, that coauthors of this community are likely to be acquainted. This finding was confirmed by an analysis of the community structure of the coauthorship and acquaintanceship networks which revealed overlapping configurations: tight-knit circles of coauthors overlap with social circles of acquaintances. This configuration, in turn, sheds light on social substructure of coauthoring events. Scientific collaboration nurtures via small-scale, localized, inter-personal networks. This finding is in agreement with previous research, mostly qualitative in nature, which suggests that collaborative synergies in cyberinfrastructure research are often the result of assemblages of people found through personal networks rather than the outcome of organizational planning and structuring \citep{nardi,knots}.

This result---that coauthors are likely to know each other---is also important for it indicates that in the context of medium-sized, inter-institutional, distributed collaborations such as the one analyzed here, acquaintanceship ties among coauthors can be safely inferred. This result, however, has to be combined with the previous finding: that acquaintanceship relations are far more numerous than coauthorship ties. Thus, while it is safe to assume that in modern cyberinfrastructure initiatives coauthoring efforts are largely based on an interpersonal bond of acquaintanceship, coauthorship cannot be utilized as an absolute proxy for acquaintanceship: acquaintanceship circles extend well beyond coauthorship circles. 

It is worth discussing how the findings presented in this article are generalizable to other scientific communities. In sum, findings of this research portray a configuration of scientific collaboration driven by social cohesion: interpersonal knowledge is the glue that holds together the scientific community under study. While these results are specific to the CENS environment, they can safely be extended to other cyberinfrastructure initiatives that are \textit{a)} similar in size and \textit{b)} similar in their multi-disciplinary component. The size of the research center is an obvious factor to consider when extending these findings to other environments. Larger, as well as smaller, research centers will naturally function on different collaboration paradigms. But the combination of social and scholarly assemblages found in a medium-sized center of few hundred researchers is likely to resemble that of CENS. A further factor to consider is the multi-disciplinary constitution of CENS. Naturally, different scientific communities function according to different authorship norms. But, being intrinsically multi-disciplinary, CENS is a hybrid milieu, where discipline-specific norms converge and blend with one another. Thus, findings of this research can be extended to medium-sized cyberinfrastructure initiatives with a multi-disciplinary orientation. 

Framing this research in the context of cyberinfrastructure studies, an important result that emerges is that social cohesion is at the basis of scientific collaboration. That is, even though the collaboratory is by definition predicated upon notions of remote collaboration and computer-supported communication, this study shows that social acquaintanceship relations are at the core of collaboration activity. Given the crucial role of social relationships for the advancement of scholarly collaboration, one wonders whether the cyberinfrastructure vision of a fluid, distributed, multi-sited science, agnostic to geographical and physical constraints, can ever be attained. In this context, this work reinforces previous recommendations to consider the spatial, social, and human arrangements that drive scientific advancement and collaboration, and how they differ across different disciplines and organizational settings \citep{olson,cummings}. Bringing this recommendation to the attention of policy makers and funding agencies has the potential to shape the direction and form of future investments and efforts in cyberinfrastructure.

\section{Acknowledgement}
This article is an abridged version of a chapter of my doctoral dissertation titled ``Structure and evolution of scientific collaboration networks in a modern research collaboratory'' defended and filed at the University of California, Los Angeles in 2010. The full text of the dissertation is openly available at \url{http://papers.ssrn.com/sol3/papers.cfm?abstract_id=1616935}. Supplemental material and replication data for the dissertation and this article, including source code and collected data, a also openly available at \url{http://hdl.handle.net/1902.1/15254}. For the preparation of this article, I would like to thank my doctoral chair Christine Borgman, as well as Mark Hansen, Marko Rodriguez, and all my former colleagues at CENS.

\bibliographystyle{elsarticle-harv}

\newpage
\appendix
\section{Acquaintanceship survey instrument}

\begin{figure}[h!]
\centering
\includegraphics[width=0.75\textwidth]{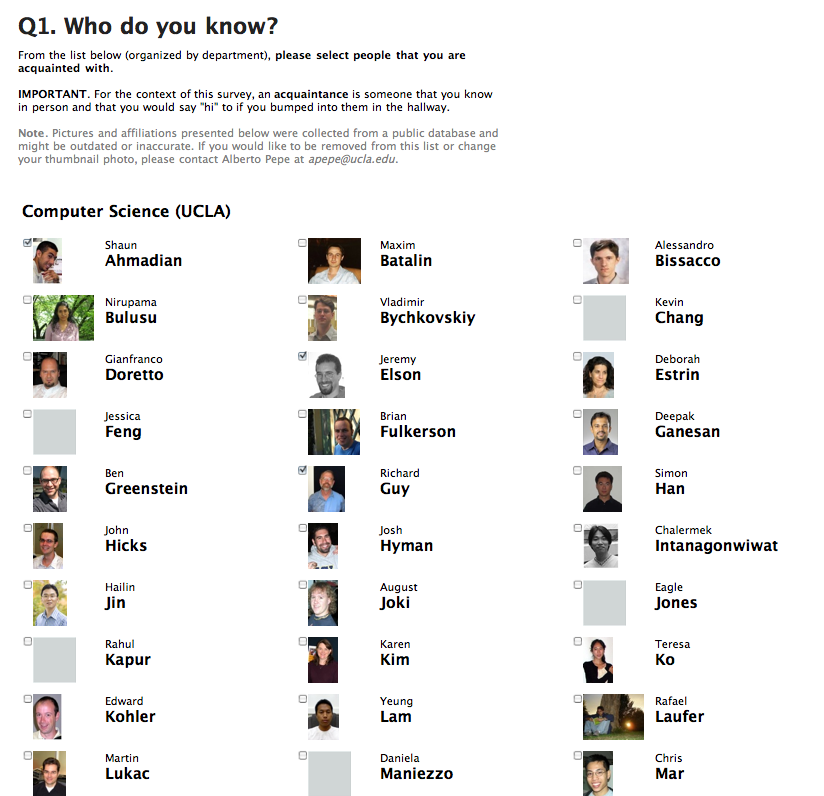}
\caption{Screenshot of the first page of the survey instrument.}
\label{fig:q1}
\end{figure}

\begin{figure}[h!]
\centering
\includegraphics[width=0.75\textwidth]{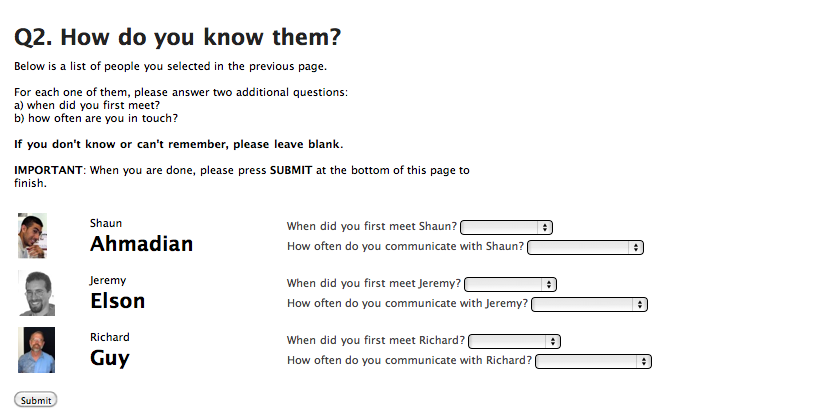}
\caption{Screenshot of the second page of the survey instrument.}
\label{fig:q2}
\end{figure}

\end{document}